

\font\twelverm=cmr10 scaled 1200    \font\twelvei=cmmi10 scaled 1200
\font\twelvesy=cmsy10 scaled 1200   \font\twelveex=cmex10 scaled 1200
\font\twelvebf=cmbx10 scaled 1200   \font\twelvesl=cmsl10 scaled 1200
\font\twelvett=cmtt10 scaled 1200   \font\twelveit=cmti10 scaled 1200

\skewchar\twelvei='177   \skewchar\twelvesy='60


\def\twelvepoint{\normalbaselineskip=12.4pt
  \abovedisplayskip 12.4pt plus 3pt minus 9pt
  \belowdisplayskip 12.4pt plus 3pt minus 9pt
  \abovedisplayshortskip 0pt plus 3pt
  \belowdisplayshortskip 7.2pt plus 3pt minus 4pt
  \smallskipamount=3.6pt plus1.2pt minus1.2pt
  \medskipamount=7.2pt plus2.4pt minus2.4pt
  \bigskipamount=14.4pt plus4.8pt minus4.8pt
  \def\rm{\fam0\twelverm}          \def\it{\fam\itfam\twelveit}%
  \def\sl{\fam\slfam\twelvesl}     \def\bf{\fam\bffam\twelvebf}%
  \def\mit{\fam 1}                 \def\cal{\fam 2}%
  \def\tt{\twelvett}
  \def\nullspace{\nulldelimiterspace=0pt \mathsurround=0pt }
  \def\big##1{{\hbox{$\left##1\vbox to 10.2pt{}\right.\nullspace$}}}
  \def\Big##1{{\hbox{$\left##1\vbox to 13.8pt{}\right.\nullspace$}}}
  \def\bigg##1{{\hbox{$\left##1\vbox to 17.4pt{}\right.\nullspace$}}}
  \def\Bigg##1{{\hbox{$\left##1\vbox to 21.0pt{}\right.\nullspace$}}}
  \textfont0=\twelverm   \scriptfont0=\tenrm   \scriptscriptfont0=\sevenrm
  \textfont1=\twelvei    \scriptfont1=\teni    \scriptscriptfont1=\seveni
  \textfont2=\twelvesy   \scriptfont2=\tensy   \scriptscriptfont2=\sevensy
  \textfont3=\twelveex   \scriptfont3=\twelveex  \scriptscriptfont3=\twelveex
  \textfont\itfam=\twelveit
  \textfont\slfam=\twelvesl
  \textfont\bffam=\twelvebf \scriptfont\bffam=\tenbf
  \scriptscriptfont\bffam=\sevenbf
  \normalbaselines\rm}



\def\beginlinemode{\endmode
  \begingroup\parskip=0pt \obeylines\def\\{\par}\def\endmode{\par\endgroup}}
\def\beginparmode{\endmode
  \begingroup \def\endmode{\par\endgroup}}
\let\endmode=\par
{\obeylines\gdef\
{}}
\def\singlespace{\baselineskip=\normalbaselineskip}

\def\doublespace{\baselineskip=\normalbaselineskip \multiply\baselineskip by 2}

\newcount\firstpageno
\firstpageno=-500

\footline={\ifnum\pageno<\firstpageno{\hfil}\else{\hfil\twelverm\folio\hfil}\fi}
\let\rawfootnote=\footnote		
\def\footnote#1#2{{\rm\singlespace\parindent=0pt\rawfootnote{#1}{#2}}}
\def\raggedcenter{\leftskip=4em plus 12em \rightskip=\leftskip
  \parindent=0pt \parfillskip=0pt \spaceskip=.3333em \xspaceskip=.5em
  \pretolerance=9999 \tolerance=9999
  \hyphenpenalty=9999 \exhyphenpenalty=9999 }


\hsize=6.5truein
\vsize=8.9truein


\parskip=\medskipamount
\twelvepoint		
\doublespace		
\overfullrule=0pt	


\def\s{\scriptscriptstyle}

\def\title			
  {\null\vskip 3pt plus 0.2fill
   \beginlinemode \doublespace \raggedcenter \bf}

\def\author			
  {\vskip 3pt plus 0.2fill \beginlinemode
   \singlespace \raggedcenter}

\def\endtitlepage		
  {\endpage			
   \body}

\def\body			
  {\beginparmode}		

\def\head#1{			
  \filbreak\vskip 0.5truein	
  {\immediate\write16{#1}
   \raggedcenter \uppercase{#1}\par}
   \nobreak\vskip 0.25truein\nobreak}

\def\refto#1{$^{#1}$}		

\def\references			
  {\head{References}		
   \beginparmode
   \frenchspacing \parindent=0pt \leftskip=1truecm
   \parskip=8pt plus 3pt \everypar{\hangindent=\parindent}}

\gdef\refis#1{\indent\hbox to 0pt{\hss#1.~}}	

\gdef\journal#1, #2, #3, 1#4#5#6{		
    {\sl #1~}{\bf #2}, #3, (1#4#5#6)}		

\gdef\journ2 #1, #2, #3, 1#4#5#6{		
    {\sl #1~}{\bf #2}: #3, (1#4#5#6)}		

\def\refstylenp{		
  \gdef\refto##1{ [##1]}				
  \gdef\refis##1{\indent\hbox to 0pt{\hss##1)~}}	
  \gdef\journal##1, ##2, ##3, ##4 {			
     {\sl ##1~}{\bf ##2~}(##3) ##4 }}

\def\refstyleprnp{		
  \gdef\refto##1{ [##1]}				
  \gdef\refis##1{\indent\hbox to 0pt{\hss##1)~}}	
  \gdef\journal##1, ##2, ##3, 1##4##5##6{		
    {\sl ##1~}{\bf ##2~}(1##4##5##6) ##3}}

\def\figurecaptions		
  {\endpage
   \beginparmode
   \head{Figure Captions}
}

\def\endpage			
  {\vfill\eject}

\def\endpaper			
  {\endmode\vfill\supereject}

\def\endit
  {\endpaper\end}


\def\frac#1#2{{\textstyle #1 \over \textstyle #2}}

\def\sla{\raise.15ex\hbox{$/$}\kern-.57em}
\def\leaderfill{\leaders\hbox to 1em{\hss.\hss}\hfill}
\def\twiddle{\lower.9ex\rlap{$\kern-.1em\scriptstyle\sim$}}
\def\bigtwiddle{\lower1.ex\rlap{$\sim$}}
\def\gtwid{\mathrel{\raise.3ex\hbox{$>$\kern-.75em\lower1ex\hbox{$\sim$}}}}
\def\ltwid{\mathrel{\raise.3ex\hbox{$<$\kern-.75em\lower1ex\hbox{$\sim$}}}}
\def\square{\kern1pt\vbox{\hrule height 1.2pt\hbox{\vrule width 1.2pt\hskip 3pt
   \vbox{\vskip 6pt}\hskip 3pt\vrule width 0.6pt}\hrule height 0.6pt}\kern1pt}

\pageno=1

%
%

\rightline{Submitted to Phys. Rev. B}
\title

Bound on the Group Velocity of an Electron in a
One-Dimensional Periodic Potential

\vskip 0.5in
\author

Michael R. Geller and Giovanni Vignale

\vskip 0.5in

{\sl Institute for Theoretical Physics
University of California
Santa Barbara, California 93016}

\vskip 0.10in
and
\vskip 0.10in

{\sl Department of Physics$^*$
University of Missouri
Columbia, Missouri 65211}

\vskip 0.5in

\body

By using a recently derived upper bound on the allowed equilibrium current in a
ring,
it is proved that the magnitude of the group velocity of a Bloch electron in a
one-dimensional periodic potential is always less than or equal to the group
velocity
of the same Bloch state in an empty lattice. Our inequality also implies that
each
energy band in a one-dimensional crystal always lies below the corresponding
free-electron band, when the minima of those bands are aligned.

\vskip 0.1in

\noindent PACS numbers: 71.25.Cx, 71.10.+x, 72.10.Bg, 73.20.Dx

\endtitlepage

\body


In a recent study,\refto{1} one of us has shown that the equilibrium orbital
current $I$
flowing through a cross section of a toroidal body cannot exceed the
fundamental limit
$$ I_{\rm max} = {N e \hbar \over 4 \pi m r_{\s 0}^2 },
\eqno(1)$$
where $N$ is the total number of electrons, $m$ is the free-electron mass, $e$
is
the magnitude of the electron charge, and
$$ {1 \over r_{\s 0}^2} \equiv {1 \over N} \int d^3r \ {n({\bf r}) \over
r_{\s \perp}^2 }
\eqno(2)$$
is the average inverse-square distance of the electrons from an axis threading
the ring.
This result is valid for general interacting electron systems in the presence
of
static but otherwise arbitrary electric and magnetic fields.\refto{2} In
particular,
it is valid for noninteracting electrons in a periodic potential, and we shall
show
that this fact leads to an unexpected---and hitherto unnoticed---constraint on
the band
structure of a one-dimensional crystal. Specifically, we shall prove here that
the
magnitude of the velocity of a Bloch electron in the $n$th band of a
one-dimensional
periodic potential is always less than or equal to the velocity of the same
Bloch state
in an empty lattice approximation. Our inequality also implies that each energy
band in
a one-dimensional crystal lies below the corresponding free-electron band, when
the
minima of those bands are aligned.

Let $E_n(k)$ be the Bloch energy levels of an electron in a one-dimensional
periodic
potential of period $a$, where $k$ is the crystal momentum. We shall label the
energy
bands according to $n = 0,1,2,\cdots,$ with $E_0(k) \leq E_1(k) \leq E_2(k)
\leq \cdots$,
so that $n$ bands are completely filled when the Fermi energy is in band $n$.
The group
velocity of an electron in band $n$ is defined as
$$ v_n(k) \equiv {1 \over \hbar} {d E_n(k) \over dk} ,
\eqno(3)$$
which is also equal to the expectation value of the velocity operator $p/m$ in
the
Bloch state $\varphi_{nk}$. Reality of the Hamiltonian implies $E_n(k) =
E_n(-k)$
and $v_n(k) = -v_n(-k)$. The existence of at most two linearly independent
solutions
of the Schr\"odinger equation at a given energy implies that the bands cannot
overlap
(however, they may touch, either at $k=0$ or at $k={\pi \over a}$, as happens
in the
free electron limit). We shall for convenience define $k_n$ to be the crystal
momentum
at the minimum of energy band $n$,
$$ k_n \equiv \bigg\lbrace \matrix{0 \ \ \ n \ {\rm even} \cr
{\pi \over a} \ \ \ n \ {\rm odd} } .
\eqno(4)$$
For a band structure with nonvanishing band gaps (that is, with energy bands
that
do not touch), the velocity clearly vanishes at $k_n$.

An upper bound on the magnitude of the group velocity may be obtained from the
inequality
$I \leq I_{\rm max}$ as follows. Let us imagine that our one-dimensional
crystal is
folded into a ring of circumference $L$ having $L/a$ unit cells. A magnetic
flux $\phi$
is now threaded through the center of the ring. This flux induces an
equilibrium persistent
current\refto{4} $I(\phi)$, which is a periodic function of flux with period
$\phi_{\s 0} \equiv hc/e$. It is easy to verify that only the partially
occupied
band---the conduction band $n$---contributes to the persistent current, and
that the
magnitude of the {\it maximum} persistent current is equal to $g e v_{\s F} /
L$,
where $v_{\s F} \equiv \big| v_n(k_{\s F}) \big|$ is the magnitude of the Fermi
velocity,
and $g=2$ is a spin-degeneracy factor. The band index $n$ of the partially
occupied band,
and the Fermi wavenumber $k_{\s F}$, may be varied to assume any desired value,
by changing
the number of electrons. Now, according to the inequality derived in Ref. 1,
$$ {g e v_{\s F} \over L} \leq I_{\rm max}.
\eqno(5)$$
Because there are $n$ comptetely filled bands and one partially filled band,
the total
number of electrons is
$$ N = gn \bigg( {L \over a}\bigg) + {gL \over \pi} \big|k_{\s F} - k_n \big|,
 \eqno(6)$$
where we have assumed that $0 \leq k_{\s F} \leq {\pi \over a}$. Substituting
this
expression into (1), and using $L = 2 \pi r_{\s 0}$, leads to our principal
result
$$ \big| v_n(k) \big| \leq {\hbar \over m} \bigg( n {\pi \over a}
+ \bigg| \big| k \big| - k_n \bigg| \bigg),
\eqno(7)$$
valid for $-{\pi \over a} \leq k \leq {\pi \over a}$.

To understand the inequality (7), we shall recall the ``band structure'' of
free
electrons, also known as the empty lattice approximation. The energy bands in
the
empty lattice are given by\refto{3}
$$ E_n^0(k) \equiv {\hbar^2 \over 2 m} \bigg(n {\pi \over a}
+ \bigg| \big| k \big| - k_n \bigg| \bigg)^{\! 2},
\eqno(8)$$
where $-{\pi \over a} \leq k \leq {\pi \over a}$. The magnitude of the group
velocity, $v_n^0(k),$ in the empty lattice, is equal to
$$ \big| v_n^0(k) \big| = {\hbar \over m} \bigg(n {\pi \over a}
+ \bigg| \big| k \big| - k_n \bigg| \bigg).
\eqno(9)$$
Therefore, our inequality (7) says that the magnitude of the velocity of a
Bloch electron in one dimension is less than or equal to the magnitude of the
velocity of the same state in the empty lattice approximation,
$$ \big| v_n(k) \big| \leq \big| v_n^0(k) \big|.
\eqno(10)$$

Integrating the inequality (7) with respect to $k$ leads to a bound on the
energy
bands themselves,
$$ E_n(k) - E_n(k_n) \leq E_n^0(k) - E_n^0(k_n) .
\eqno(11)$$
Therefore, each energy band must lie below the corresponding free-electron band
when the minima of those bands are aligned. In particular, the actual band
widths
$W_n$ are bounded by the free-electron band widths
$$ W_n^0 = (2n+1) {\hbar^2 (\pi / a)^2 \over 2 m} .
\eqno(12)$$

As a final comment, we note that it is possible to prove the inequality (7)
directly. We start with the well-known identity
$$ {1 \over \hbar^2} {d^2 E_n \over d k^2} = {1 \over m} - {1 \over m}
\sum_{n' \neq n} f_{nn'}(k),
\eqno(13)$$
where the
$$ f_{nn'}(k) \equiv {2 \over m} \ {|p_{nn'}(k)|^2 \over E_{n'}(k)
- E_n(k) }
\eqno(14)$$
are the usual oscillator strengths. Here the $p_{nn'}(k) \equiv \big\langle
\varphi_{nk} \big| p \big| \varphi_{nk} \big\rangle$ are matrix elements of the
momentum operator between Bloch states with the same crystal momentum. The
oscillator strengths have the properties $f_{nn'}(k) \geq 0$ for $n' > n$, and
$f_{nn'}(k) = - f_{n' n}(k)$. Note that
$$ \int_0^{\pi \over a} dk \ {d^2 E_n \over dk^2} = 0,
\eqno(15)$$
so that
$$ \int_0^{\pi \over a} dk \ \sum_{n' \neq n} f_{nn'}(k) = {\pi \over a},
\eqno(16)$$
for any band $n$. Also note that
$$ \sum_{n' = 0}^n \sum_{n'' \neq n'} f_{n' n''}(k) \geq 0,
\eqno(17)$$
because the contribution from terms with $n'' < n$ vanishes, leaving only
positive-definite oscillator strengths. This inequality implies that
$$ \sum_{n' \neq n} f_{nn'}(k) \geq - \sum_{n' = 0}^{n-1}
\sum_{n'' \neq n'} f_{n' n''}(k) .
\eqno(18)$$
Next, we integrate  (13) from $0$ to $k \geq 0$ and use the inequality (18)
to obtain
$$ {1 \over \hbar^2} {d E_n \over dk} \leq {k \over m} + {1 \over m} \int_0^k
dk'
\ \sum_{n' = 0}^{n-1} \sum_{n'' \neq n'} f_{n' n''}(k').
\eqno(19)$$
According to (17), the integrand on the right-hand-side of (19) is
positive-definite, so
$$ {1 \over \hbar^2} {d E_n \over dk} \leq {k \over m} + {1 \over m}
\int_0^{\pi \over a}
dk' \ \sum_{n' = 0}^{n-1} \sum_{n'' \neq n'} f_{n' n''}(k'),
\eqno(20)$$
or, upon using (16),
$$ {1 \over \hbar^2} {d E_n \over dk} \leq {1 \over m} \bigg( n {\pi \over a}
+ k \bigg) ,
\eqno(21)$$
valid for $0 \leq k \leq {\pi \over a}$. The upper bound (21) is valid for any
band
$n$, but it is not useful for odd $n$, which have negative velocities for
$0 \leq k \leq {\pi \over a}$. However, we may also obtain a lower bound on the
velocity by integrating (13) from $k$ to ${\pi \over a}$ and using (18), which
leads to
$$ {1 \over \hbar^2} {d E_n \over dk} \geq {1 \over m} \bigg( k-(n+1) {\pi
\over a}
\bigg) .
\eqno(22)$$
For odd $n$ and $0 \leq k \leq {\pi \over a}$, this is equivalent to
$$ {1 \over \hbar^2} \bigg| {d E_n \over dk} \bigg| \leq {1 \over m} \bigg(
(n+1) {\pi \over a}
- k \bigg) .
\eqno(23)$$
The inequalities (21) and (23) together prove our upper bound (7) directly.

This work was supported by the National Science Foundation through Grant
Numbers
DMR-9100988 and PHY89-04035.

\references

\refto{*}Permanent address.

\refto{1} G. Vignale (to be published).

\refto{2} We note here that (1) is modified by current-current interactions, an
order
$v^2/c^2$ relativistic effect of the magnetic field produced at a point ${\bf
r}$
by a moving electron at ${\bf r}'$.

\refto{3} The reduced-zone-scheme energy bands in the empty lattice
approximation
are obtained by folding the free electron dispersion relation into the first
Brillouin zone of the actual crystal.

\refto{4} M. B\"uttiker, Y. Imry, and R. Landauer, Phys. Lett. {\bf 96A}, 365
(1983).

\endit